\documentstyle[pra,aps,twocolumn,floats,psfig]{revtex}

\begin{document} \title{Bose-Einstein condensation in
variable dimensionality} \author{B. A. McKinney\thanks{e-mail:
brett\_mckinney@mailaps.org} and D. K. Watson\thanks{e-mail:
dwatson@ou.edu}}
\address{Department of Physics and Astronomy, University of
Oklahoma, Norman, OK 73019}

\date{\today}
\maketitle

\begin{abstract}
We introduce dimensional perturbation techniques to Bose-Einstein
condensation of inhomogeneous alkali gases (BEC).  The
perturbation parameter is $\delta=1/\kappa$, where $\kappa$
depends on the effective dimensionality of the condensate and on
the angular momentum quantum number. We derive a simple
approximation that is more accurate and flexible than the $N
\rightarrow \infty$ Thomas-Fermi ground state approximation (TFA)
of the Gross-Pitaevskii equation.  The approximation presented
here is well-suited for calculating properties of states in three
dimensions and in low effective dimensionality, such as vortex
states in a highly anisotropic trap.
\end{abstract}

\section {Introduction}

The most commonly used approach to describe a dilute gas of atoms
in a BEC at $T=0$ is mean-field theory, which takes the form of
the time-independent Gross-Pitaevskii equation (GPE):
\begin{eqnarray}
{\displaystyle -{\hbar^2 \over 2m} \bigtriangledown^2\psi({\bf r})
+ V_{trap}(r) \psi({\bf r}) + N U_{3} {|\psi({\bf r})|}^2
\psi({\bf r})} & \nonumber \\
{\displaystyle =\mu \psi({\mathbf r}),} \label{GPE}
\end{eqnarray}
where the three-dimensional coupling constant is $U_{3}=(4\pi
\hbar^2 a)/m$, a is the s-wave scattering length, $V_{trap}$ is
the external trapping potential, $\mu$ is the chemical potential,
and N is the number of condensate atoms.  The complex order
parameter, $\psi({\bf r})$, is referred to as ``the wave function
of the condensate.''

The $N \rightarrow \infty$ Thomas-Fermi approximation (TFA) has
been proven to be a highly successful analytical approximation of
the GPE \cite{baym,edwards}.  The strength of the $N \to \infty$
TFA is its simplicity: neglecting the kinetic energy results in a
simple approximation of the ground-state condensate density that
is effective in analyzing properties of large-N condensates. For
condensates with a moderate number of atoms and condensates with
attractive interactions, the TFA  breaks down. The kinetic energy
is important in each case, especially in the latter, where the
kinetic energy is necessary to prevent collapse.  The effects of
attractive interactions have been studied using approximation
techniques such as variational trial wave functions
\cite{stoof,fetvar,perez}. Other approximations that have been
employed to extend the TF regime of validity include the $\hbar
\rightarrow 0$ TFA \cite{schuck}, a method that uses two-point
Pad\'{e} approximants between the weakly- and strongly-interacting
limits of the ground-state \cite{gonzalez}, and a variational
method for anisotropic condensates \cite{das}.

With the study of BEC in highly anisotropic traps as a motivation,
we use a perturbation formalism that permits the effective
dimensionality, D, to vary. Because it readily allows one to
approximate quantities in any dimension, such a formalism is
ideally suited for condensates in the anisotropic traps used in
many laboratories where the condensate can be effectively one-,
two- or three-dimensional. The perturbation parameter is
$\delta=1/\kappa$, where $\kappa$ depends on the effective
dimensionality of the condensate and on the angular momentum
quantum number.  The $\delta \to 0$ limit becomes an exactly
soluble problem, the solution of which is used by the various
dimensional-scaling methods as the starting point for the solution
of the full three-dimensional problem \cite{Copen,note1}. The
$\delta \to 0$ approximation to the condensate density, which
retains part of the kinetic energy, is quite accurate for both a
large and moderate number of atoms in the BEC ground state, and
the dimensional scaling formalism, which treats the dimensionality
as a parameter, is advantageous when studying condensates of low
effective dimensionality due to extreme trap anisotropy. The
centrifugal term in the $\delta \to 0$ density also makes it a
good physical starting point for treating vortex states.

\section {$N \rightarrow \infty$ Thomas-Fermi Approximation to the Ground-state}

In the case of positive scattering length, the repulsive
interaction causes the density to become flat, and the kinetic
energy of the condensate becomes negligible in the $N \rightarrow
\infty$ limit. This limit of the GPE results in the highly
successful classical approximation for the density of the
condensate ground-state known as the Thomas-Fermi approximation
($N \to \infty$ TFA).  The $N \to \infty$ TFA for the ground state
in a three-dimensional isotropic trap is \cite{baym,edwards}
\begin{equation}
\label{TFdensity} \rho_{TF}(r)={|\psi|}^2 = {1\over{N U_{3}}}
(\mu_{TF} - {1\over2} m \omega^2 r^2)
\end{equation}
for $\mu_{TF} \ge {1\over2} m \omega^2 r^2$ and $\rho_{TF}=0$
elsewhere.   Eq.\ (\ref{TFdensity}) provides an excellent
description of the condensate ground-state density in the bulk
interior.  This approximation breaks down near the surface of the
gas where the density is not flat; the wave function must vanish
smoothly, making the kinetic energy appreciable in the boundary
layer.  The chemical potential is obtained from continuity and
normalization of Eq.\ (\ref{TFdensity}).  In oscillator units, one
finds
\begin{equation}
\mu_{TF}={1\over2} R^2,
\end{equation}
where $R=(15 N a/a_{ho})^{1/5}$ is the Thomas-Fermi classical
cutoff radius in oscillator units of length, $a_{ho}=\sqrt{\hbar/m
\omega}$.

Boundary layer theory techniques have been employed to obtain
corrections to the $N \to \infty$ TFA at the condensate surface
where the gradient of the density is no longer small
\cite{dalfovo,lundh,fetfed}.  The leading order correction to the
ground state chemical potential due to the boundary layer at the
surface is of order $R^{-4}ln(R)$.

\section {Effective Dimensionality}

We use a perturbation formalism where the effective
dimensionality, D, of the condensate is allowed to vary.  The
effective-dimensionality of the condensate depends on the relative
size of the confinement length in each of the three spatial
dimensions.  Most experimentally realized traps are axially
symmetric with some having a high degree of anisotropy
\cite{mit,jila}.  In the case of axial symmetry, the trapping
potential takes the form, $V_{trap}(r) = {1 \over 2} m
\omega_{\perp}^2 r_{\perp}^2 + {1 \over 2} m \omega_{z}^2 z^2 = {1
\over 2} m \omega_{\perp}^2 (r_{\perp}^2 + \lambda^2 z^2)$, where
$\lambda=\omega_{z}/\omega_{\perp}$ is a measure of the degree of
anisotropy.  The system reduces to a three-dimensional isotropic
condensate for $\lambda = 1$.  In the small- (large-) $\lambda$
limits, the system reduces to an effective one- (two-)dimensional
isotropic condensate.  It is conceivable that an isotropic
hamiltonian in  a fractional-dimensional space could be used to
describe experimental condensates for intervening values of
$\lambda$, but we will focus our attention on integer dimensions.

As an illustration, consider $\lambda \gg 1$ where the motion of
the atoms in the z-direction becomes frozen and their motion is
described by a gaussian of small width.  To determine the 2D
effective coupling constant, we assume the wave function in Eq.\
(\ref{GPE}) is separable: $\psi({\bf r}) = \psi_{2}(r_{\perp})
\chi(z)$, where $\chi(z)$ is assumed to be a gaussian, and
operating with $\int dz \chi^{\ast}$, one finds a new effective 2D
GPE:
\begin{equation}
\label{2dex} \biggl(-{\hbar^2 \over 2m} \bigtriangledown_{2}^2 +
{1 \over 2} m \omega_{2}^2 r^2 + N_{2} U_{3} {|\psi_{2}|}^2
\biggr) \psi_{2}=\mu_{2} \psi_{2},
\end{equation}
which has the same form as Eq.\ (\ref{GPE}), but r is the 2D
radius, $\omega_{2}=\omega_{\perp}$, $\mu_{2}=\mu - \hbar
\omega_{z}/2$, $N_{2} = N  \int dz |\chi|^4$. Requiring that
$\psi_{2}$ and $\chi$ be normalized to unity, $\int dz |\chi|^4$
has units of 1/length; thus, we interpret $N_{2}$ as the number of
atoms in the 2D condensate per unit length along the z-axis.  In
our subsequent scalings, we will adopt a notation for the number
of atoms that is similar to that of Jackson et. al \cite{jackson}.
For $\lambda \ll 1$, one may assume the motion in the
radial-direction in the x-y plane is described by a gaussian of
small width, $\chi(r_{\perp})$, and, following the same procedure,
one obtains a 1D equation analogous to Eq.\ (\ref{2dex}), where
$N_{1}$ would represent the number of atoms in the 1D condensate
per unit area in the xy-plane.

\section {GPE in variable dimensionality}
\label{sec:isotropic}

We begin by explicitly generalizing the nonlinear Schr\"odinger
equation (NLSE), Eq.\ (\ref{GPE}), to D-dimensions where r becomes
the radius of a D-dimensional sphere with D-1 remaining angles.
The Laplacian is generalized to D-dimensions (Bohn, Esry and
Greene\cite{beg,bec1} treat the Laplacian in a similar fashion, in
which they use hyperspherical coordinates to define a mean
condensate radius.), and the potential terms retain their
three-dimensional form; the coupling constant is generalized in
the final scaling.  We obtain the Schr\"odinger equation:
\begin{eqnarray}
{\displaystyle \Biggl\{-{\hbar^2 \over 2m} \Bigl[{1\over r^{D-1}}
{\partial \over
\partial r} \Bigl({r^{D-1}} {\partial \over \partial r}\Bigr) +
{L_{D-1}^2 \over r^2}\Bigr] + {1\over2} m {\omega}^2 r^2} &
\nonumber \\ {\displaystyle + N U_{3} {|\psi({\bf r})|}^2\Biggr\}
\psi({\bf r}) = \mu \psi({\bf r}),}
\end{eqnarray}
where $L_{D-1}^2$ is a generalized angular momentum operator
depending on D-1 angles with eigenvalues $-l(D + l - 2)$
\cite{avery}; the angular momentum quantum number, $l$, is
non-negative. Substituting these eigenvalues and introducing the
radial Jacobian factor in a transformation of the wave function,
$\phi(r) = r^{(D-1)/2} \psi(r)$, to eliminate the first derivative
terms, we find
\begin{eqnarray}
{\displaystyle \Biggl\{-{\hbar^2 \over 2m} \Bigl[{\partial^2 \over
\partial r^2} -{{(D - 1)(D - 3)} \over {4 r^2}} - {{l (D + l
-2)} \over r^2}\Bigr]} & \nonumber \\ {\displaystyle + {1\over2} m
{\omega}^2 r^2 + N U_{3} {|\psi({\bf r})|}^2\Biggr\} \phi({\bf r})
= \mu \phi({\bf r}).}
\end{eqnarray}

Finally, we make two sets of scalings to arrive at the NLSE in
dimensionally scaled oscillator units. The first scaling is a
purely dimensional scaling: $r=\kappa^2 \tilde r$,
$\tilde{\omega}=\kappa^3 \omega$, $\tilde{\mu}=\kappa^2 \mu$, and
$\tilde{\psi}=\kappa^D \psi$, where $\kappa = D + 2 l$. The final
scaling is to scaled oscillator units (denoted by bars):
$\tilde{r}=\tilde{a}_{ho} \bar{r}$, $\tilde{\mu}=\hbar
\tilde{\omega} \bar{\mu}$, and $\bar{\psi}=\tilde{a}^{D/2}_{ho}
\tilde{\psi}$, where
$\tilde{a}_{ho}=\sqrt{\hbar/m\tilde{\omega}}$.  Combining these
two scalings, we arrive at
\begin{eqnarray}
{\displaystyle \Biggl\{- {1\over2}\delta^2 {\partial^2 \over
\partial \bar{r}^2} + {{1 - 4\delta + 3\delta^2} \over {8
\bar{r}^2}}   + {1\over 2} \bar{r}^2} & \nonumber \\
{\displaystyle + \bar{g}_{D} {|\bar{\psi}(\bar{{\bf r}})|}^2
\Biggr\} \bar{\phi}(\bar{\bf r}) = \bar{\mu} \bar{\phi}(\bar{\bf
r}),} \label{scaledGPE}
\end{eqnarray}
where everything is now in dimensionally scaled oscillator units
and $\delta = 1/\kappa$.  For the effective dimensions of primary
interest in this study, the dimensionally scaled coupling
constants are, for 3D, $\bar{g}_{3} = g_{3}/\kappa^{5/2}$, where
$g_{3}=4 \pi N_{3} a/a_{ho}$ and $N_{3}$ is the number of
condensate atoms; and for 2D, $\bar{g}_{2} = g_{2}/\kappa^{2}$,
where $g_{2}=4 \pi N_{2} a$ and $N_{2}$ represents the number of
atoms in the 2D condensate per unit length along the z-axis.  The
definition of $N_{2}$ makes $g_{2}$ dimensionless. For general D,
\begin{equation}
\bar{g}_{D}={{4\pi N_{D} a}\over{\kappa^{{D+2}\over2}
a^{D-2}_{ho}}},
\end{equation}
making Eq.\ (\ref{scaledGPE}) valid for describing condensates in
any effective dimension.  In the next section, we describe a
simple and accurate zeroth-order approximation to Eq.\
(\ref{scaledGPE}).

\section {zeroth-order density}

It has been pointed out by Schuck and Vi\~{n}as \cite{schuck} that
the true TF limit ($\hbar \rightarrow 0$ as originally applied to
the case of Fermi statistics \cite{thomas}) is not equivalent to
$N \rightarrow \infty$, and they show that the $\hbar \rightarrow
0$ TF limit for bosons does not neglect the kinetic energy for the
ground state. The $N \to \infty$ TFA to the ground state is too
harsh on the kinetic energy for a moderate number of atoms. A less
harsh and nearly as simple approximation is the zeroth-order
($\delta \to 0$) approximation of Eq.\ (\ref{scaledGPE}).  Unlike
the ground-state $N \to \infty$ TFA, which neglects the entire
kinetic energy, our zeroth-order approximation of the generalized
GPE neglects the derivative part of the kinetic energy but retains
a centrifugal term. For vortex states, one understands this term
as being a centrifugal barrier due to quantized circulation, which
pushes atoms away from the axis of rotation. This centrifugal
barrier arises from the condensate phase: $|\nabla S|^2$ ($\psi =
\sqrt{\rho} e^{i S}$, where $\rho$ is the condensate density and
$S$ is the spatially dependent condensate phase; then the
condensate velocity is given by ${\bf v}=\frac{\hbar}{m} \nabla
S$).  For the ground state, this centrifugal term in the
zeroth-order density has an alternate, quantum mechanical
interpretation, which helps explain its good agreement with
numerical calculations.  We discuss this interpretation in Section
\ref{sec:3d}.

The $\delta \to 0$ limit of the angular-dimensional perturbation
parameter in Eq.\ (\ref{scaledGPE}) results in the following
zeroth-order density in scaled oscillator units:
\begin{equation}
\label{DPTdensity} \rho(\bar{r})={|\bar{\psi}|}^2 =
{1\over{\bar{g}_{D}}} (\bar{\mu} - {1\over{8 \bar{r}^2}} -
{1\over2} \bar{r}^2),
\end{equation}
for $\bar{R}_{o}(\bar{\mu}) \le \bar{r} \le
\bar{R}_{max}(\bar{\mu})$ and $\rho=0$ elsewhere.  The
normalization condition becomes
\begin{equation}
\label{norm} \Omega(D)
\int\limits_{\bar{R}_{o}(\bar{\mu})}^{\bar{R}_{max}(\bar{\mu})}
d\bar{r} \bar{r}^{D-1} {|\bar{\psi}|}^2 = 1,
\end{equation}
where $\Omega(D)=2 \pi^{D/2}/\Gamma(D/2)$\footnote{As the GPE is
nonlinear, one cannot treat excited-$l$ states (vortices) for
$D>2$ radially symmetric traps in the usual manner of separating
the wave function into radial and angular parts. Presently,
however, vortices in 3D radially symmetric traps are not realized.
Vortex states in a 2D isotropic trap do not pose a problem to
theory because the spherical harmonic wave function acts as a
phase factor.  In 1D, the spherical harmonic wave function is a
constant and, since there are no angles, one can think of $l=0$
and $l=1$ as even and odd parity states.}.  The $\delta \to 0$
limit can be thought of as a large-D or large-l limit.


Eq.\ (\ref{DPTdensity}) is valid where the density is
non-negative. In addition to the $N \to \infty$ TF-like classical
cutoff radius near the surface, $\bar{R}_{max}$, the centrifugal
term requires that another cutoff be defined, $\bar{R}_{o}$,
slightly removed from the origin, to satisfy the requirement that
the density be non-negative.  In terms of the chemical potential,
the cutoff radii in scaled oscillator units are defined as
\begin{eqnarray}
& \bar{R}_{o}^2(\bar{\mu}) = \bar{\mu}-\sqrt{
\bar{\mu}^2-{1\over4}} & \nonumber \\
 & and & \nonumber \\
& \bar{R}_{max}^2(\bar{\mu}) =
\bar{\mu}+\sqrt{\bar{\mu}^2-{1\over4}}. &  \label{continuity}
\end{eqnarray}
In regular oscillator units ($a_{ho}$),
\begin{eqnarray}
& R_{o}^2(\mu) = \mu-\sqrt{ \mu^2-{{\kappa^2}\over4}} & \nonumber
\\ & and & \nonumber \\
& R_{max}^2(\mu) = \mu+\sqrt{\mu^2-{{\kappa^2}\over4}}. &
\end{eqnarray}
Notice for the ground state in the strongly interacting regime
that $\mu \gg 1$ and the cutoff radii for the ground state become
$N \to \infty$ TF-like: $R_{o} \approx 0$ and $\mu \approx
R_{max}^2/2$; the strongly interacting limit, or, equivalently,
the $N\rightarrow\infty$ limit of our zeroth-order approximation
collapses to the $N \to \infty$ TFA, as expected.  (For finite N,
as will be shown later, our zeroth-order approximation gives
better agreement with the numerical solution of the GPE than the
$N \to \infty$ TFA.) Using the integration limits defined in Eq.\
(\ref{continuity}), along with the condensate density defined in
Eq.\ (\ref{DPTdensity}), the normalization condition (Eq.\
\ref{norm}) gives an equation for the zeroth-order chemical
potential that is easily solved in any dimension. (See Section
\ref{sec:lowerD} where this procedure is illustrated for two
dimensions.)

Once the chemical potential is calculated, it is then used in Eq.\
(\ref{DPTdensity}) to complete the description of the zeroth-order
wave function.  One can then calculate the energy from
\begin{eqnarray}
E/N & = & \int d^{D} {\bf r} \biggl[ {{\hbar^2}\over{2m}}|\nabla
\psi|^2 + {1\over2}m\omega^2 r^2 \psi^2 + {g_{D}\over2} \psi^4
\biggr] \nonumber \\
  & = & E_{kin}/N + E_{ho}/N + E_{int}/N,
\end{eqnarray}
or in the zeroth-order approximation and scaled oscillator units,
\begin{eqnarray}
\bar{E}/N & \approx & \Omega(D)
\int\limits_{\bar{R}_{o}(\bar{\mu})}^{\bar{R}_{max}(\bar{\mu})}
\bar{r}^{D-1} d\bar{r} \biggl[ {1\over {8 \bar{r}^2}} \bar{\psi}^2
+ {1\over2} \bar{r}^2 \bar{\psi}^2 + {{\bar{g}_{D}}\over2}
\bar{\psi}^4 \biggr] \nonumber \\
  & \approx & \bar{E}_{kin}/N +
\bar{E}_{ho}/N + \bar{E}_{int}/N. \label{energyperatom}
\end{eqnarray}

\section {Ground state in three dimensions}
\label{sec:3d}

The numerical effect of the centrifugal-like term in the
zeroth-order approximation on the ground state of a stationary
condensate is clear from Fig.\ \ref{fig:mucompare}, where we
compare the numerical solution of the GPE chemical potential with
our zeroth-order approximation and the $N \to \infty$ TFA for up
to $10,000$ $^{87}Rb$ atoms in a spherical trap. Our zeroth-order
approximation is more accurate than the $N \to \infty$ TFA for all
N, most notably for a moderate number of atoms. The accuracy of
the zeroth-order approximation is comparable to boundary layer
corrections: the zeroth-order approximation is slightly more
accurate for small coupling constant, while boundary layer theory
is slightly more accurate for larger coupling constant, but the
difference between all three approximations becomes small for very
large coupling constant.

The correct physical interpretation of this centrifugal-like term,
as originally noted by Chatterjee\cite{chat}, is that it is the
component of the kinetic energy needed to satisfy the minimum
uncertainty principle.  The zeroth-order density includes a
centrifugal term from the kinetic energy, which pushes the wave
function away from the origin in the ground state as if there were
a non-zero quantum of angular momentum; however, the $1/r^2$
contribution to the ground state density of a nonrotating cloud
clearly is not due to any rotational motion of the cloud.  This
effect, which becomes less pronounced as N increases, is
demonstrated in Figs.\ \ref{fig:gwf1} and \ref{fig:gwf2}, which
show the numerically calculated GPE ground-state non-Jacobian
weighted wave function ($\psi$) along with our zeroth-order
approximation and the $N \to \infty$ TFA.

The centrifugal-like term in the lowest order of dimensional
perturbation theory ($\delta \to 0$) can be understood as arising
from the requirement that the system's uncertainty product be a
minimum\cite{chat}. Another way to see how a centrifugal term may
arise in the ground state -- this time within the $N \to \infty$
TFA -- is by applying the Langer modification of WKB theory to the
$N \to \infty$ TF density.  For vortices, one may not neglect the
entire kinetic energy in the $N \to \infty$ limit.  A slightly
more general $N \to \infty$ TF density than Eq. (\ref{TFdensity})
that includes vortices is
\begin{equation}
\rho_{TF}(r)={|\psi|}^2 = {1\over{N U_{3}}} (\mu_{TF} -
\frac{\hbar^2 \Lambda^2}{2mr^2}- {1\over2} m \omega^2 r^2).
\end{equation}
For a spherical trap, $\Lambda^2=l(l+1)$, which reduces to the
usual ground state $N \to \infty$ TF density for $l=0$, but using
the Langer modification, where the correct asymptotic phase of the
WKB wave function is obtained by the replacement $l (l+1)
\rightarrow (l+1/2)^2$ in the centrifugal potential, a centrifugal
barrier remains in the ground-state:
\begin{equation}
\rho_{TF}(r) \to {|\psi|}^2 = {1\over{N U_{3}}} (\mu_{TF} -
\frac{\hbar^2}{8mr^2}- {1\over2} m \omega^2 r^2).
\end{equation}

The dependence of our perturbation parameter on the angular
momentum quantum number suggests that the zeroth-order density
will be a good physical starting point for vortex states, which we
explore in the next section for $D=2$. The remaining centrifugal
term in our zeroth-order approximation is a lowest-order
correction to the kinetic energy, which, for a moderate number of
atoms, greatly improves the ground state approximation ($l=0$)
over the $N \to \infty$ TFA.  For a very large number of atoms,
the contribution from the kinetic energy becomes very small, as
can be seen by comparing the wave functions in Figs.\
\ref{fig:gwf1} and \ref{fig:gwf2}, for $N=10^4$ and $N=10^6$
$^{87}Rb$ atoms, respectively. As the number of atoms increases,
our unphysical core becomes smaller than the healing length,
eventually vanishing: our zeroth-order and the $N \to \infty$ TF
wave functions become indistinguishable from the numerical
solution for large N.

\section {Lower dimension}
\label{sec:lowerD}

\begin{figure}[H]
\psfig{figure=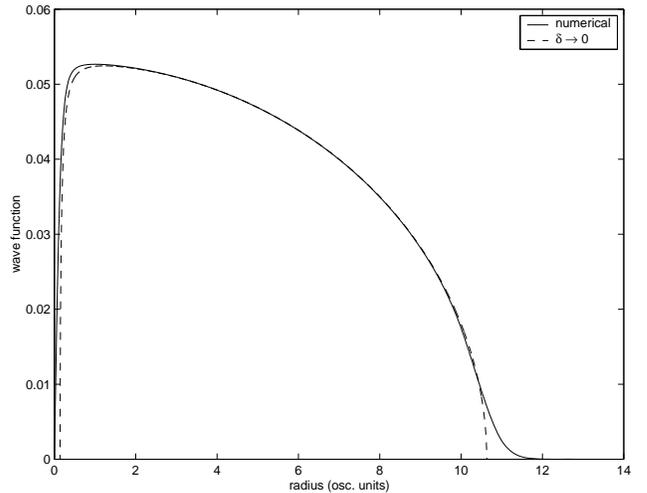,height=6.6cm,width=8.6cm}
\caption[fig:gwf1]{Comparison of the condensate wave function
($\psi$ -- non-Jacobian weighted) with an $l=1$ vortex in a 2D
isotropic trap with $g_{2}=10,000$. The solid line is the
numerical solution of the GPE and the dashed line is our
zeroth-order ($\delta \to 0$) wave function, whose centrifugal
term models the vortex core.} \label{fig:2dvort}
\end{figure}

The $\delta \to 0$ density (Eq. \ref{DPTdensity}) is well suited
for describing condensates in the presence of a vortex, where the
centrifugal term models the vortex core (see Fig.
\ref{fig:2dvort}). In this section, we present explicit
expressions for the $D=2$ ground-state and vortex states.  In the
angular-dimensional scaling of the GPE, one has considerable
freedom in the choice of the scaling parameter, $\delta
=1/\kappa$. In the previous section, we used $\kappa = D +2 l$ or,
for the ground state, $\kappa = D$.  The choice $\kappa = D + 2 l
-2$ exactly reduces our expressions below for the chemical
potential and energy to a two-dimensional $N \to \infty$ TFA that
includes the leading contribution to the kinetic energy due to
fluid motion of the condensate\cite{crescim}. Slightly improved
agreement of the zeroth-order energy with the numerical solution
of the 2D GPE for a wide range of mean-field coupling constant can
be obtained by choosing $\kappa = D + 2 l -1$, which changes the
numerator in the centrifugal term of Eq. (\ref{scaledGPE}) to
$1-\delta^2$. Zeroth order predictions show a small amount of
variability with the choice of $\kappa$, but the results of higher
order perturbation theory should not depend on the particular
choice of $\kappa$.

Using Eqs.(\ref{DPTdensity},\ref{norm},\ref{continuity}) for
$D=2$, one finds that the scaled chemical potential satisfies
\begin{equation}
\label{2dmueqn} {\bar{g}_{2} \over {2 \pi}} = {\bar{\mu}\over2}
\sqrt{ \bar{\mu}^2 -{1\over4}} + {1 \over 16} \ln \Biggl({{
\bar{\mu}-\sqrt{\bar{\mu}^2 -{1\over4}} \over {\bar{\mu}+\sqrt{
\bar{\mu}^2 -{1\over4}}}}} \Biggr).
\end{equation}
Recalling the earlier conversion relations leading to Eq.\
(\ref{scaledGPE}), the chemical potential, in regular oscillator
units, satisfies
\begin{equation}
\label{scaled2dmueqn} {g_{2}\over {2 \pi}}  = {{\mu}\over2} \sqrt{
\mu^2 -{{\kappa^2}\over4}} + {{\kappa^2} \over 16} \ln \Biggl({{
\mu-\sqrt{\mu^2 -{{\kappa^2}\over4}} \over {\mu+\sqrt{ \mu^2
-{{\kappa^2}\over4}}}}} \Biggr).
\end{equation}
Solving Eq.\ (\ref{2dmueqn}) for the scaled zeroth-order chemical
potential, $\bar{\mu}$, and using the resulting wave function,
Eq.\ (\ref{DPTdensity}) and Eq.\ (\ref{energyperatom}) give a
simple analytical approximation for the 2D energy,
\begin{equation}
\label{2dEeqn} E = {{2 \pi} \over {3 g_{2}}} \biggl(\mu^2 -
{{\kappa^2}\over4} \biggr)^{3/2},
\end{equation}
where we have already converted to regular oscillator units. Eq.\
(\ref{scaled2dmueqn}) and Eq.\ (\ref{2dEeqn}) for the chemical
potential and energy per atom, respectively, are analagous to the
results of the $N \to \infty$ TFA for vortices given in Ref.
\cite{crescim}, which includes a kinetic energy term associated
with the fluid motion that is encoded in the wave function's
phase. This similarity is due to the zeroth-order $\delta \to 0$
limit being a large angular momentum limit or, in the language of
hydrodynamics, a large quantum of circulation limit. The
zeroth-order approximation for $D=2$ results in a shifted TF-like
energy spectrum, whose ground state approximation is, just as for
$D=3$, more accurate than the $N \to \infty$ TFA for any coupling
constant, most noticeably for smaller coupling. For the energy of
a single charge vortex located at the center of the trap, the
above expressions are more accurate than the unregulated $N \to
\infty$ TFA in Ref. \cite{crescim} for a moderately sized coupling
constant, and slightly less accurate for very large coupling.  For
our zeroth-order approximation and $N \to \infty$ TFA,
respectively, the relative errors in the first vortex state energy
are $0.56\%$ and $-0.88\%$ for $g_2 = 1,000$; and $0.015\%$ and
$-0.012\%$ for $g_2 = 100,000$.

\section {Conclusions}

\begin{figure}[H]
\psfig{figure=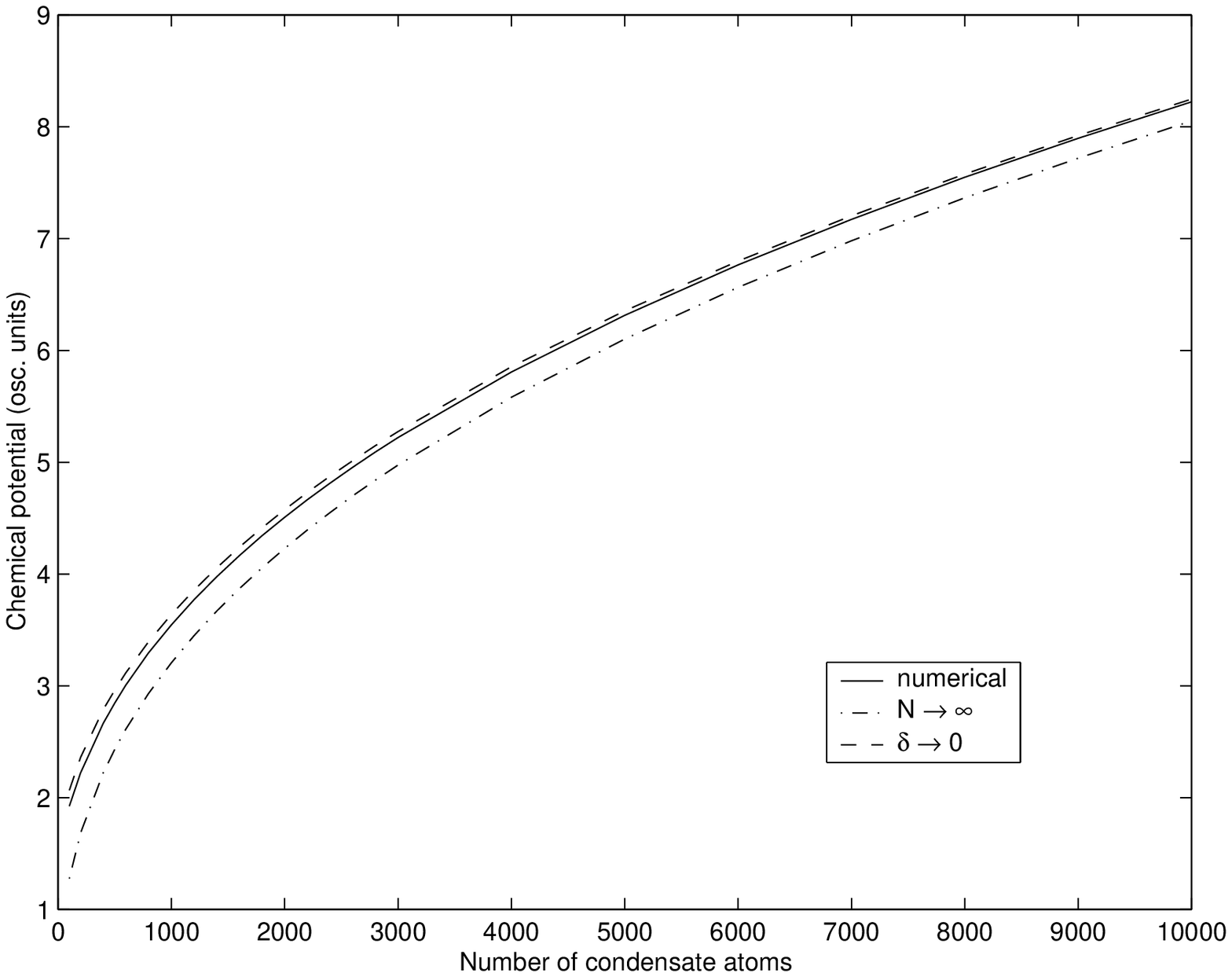,height=6.6cm,width=8.6cm}
\caption[fig:mucompare]{Chemical potential in oscillator units vs.
number of condensate atoms for a $^{87}Rb$ condensate in a 3D
isotropic trap, where $a=100$ bohr and $\nu = 200$ Hz.  The
zeroth-order ($\delta \to 0$) approximation of dimensional
perturbation theory presented here (dashed) is in better agreement
with the numerical solution of the GPE (solid) than the $N \to
\infty$ Thomas-Fermi approximation (dash-dot).}
\label{fig:mucompare}
\end{figure}

\begin{figure}[H]
\psfig{figure=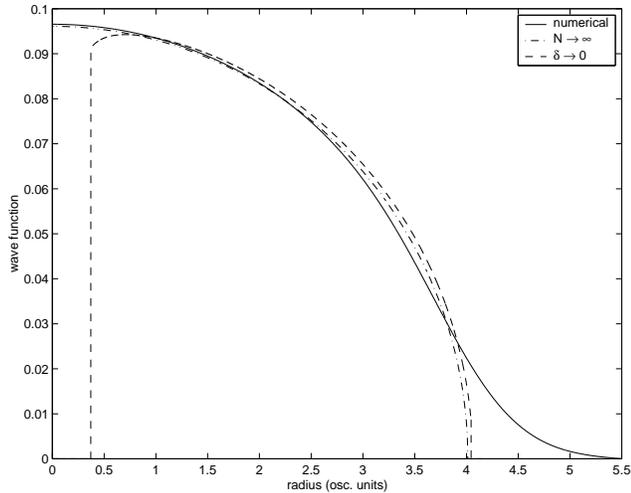,height=6.6cm,width=8.6cm}
\caption[fig:gwf1]{Ground state wave functions ($\psi$ --
non-Jacobian weighted) for a $^{87}Rb$ condensate of 10,000 atoms
in a 3D isotropic trap, where $a=100$ bohr and $\nu=200$ Hz. These
parameters correspond to $g_{3} \approx 872.04$.  Plotted are the
numerical solution of the GPE (solid), the $N \to \infty$
Thomas-Fermi approximation (dash-dot) and our zeroth-order
($\delta \to 0$) approximation (dashed).  Our zeroth-order
approximation contains an unphysical core near the origin, but the
added kinetic energy, which causes the core to appear, is also
responsible for the increased accuracy seen in Fig.\
\ref{fig:mucompare}.} \label{fig:gwf1}
\end{figure}

\begin{figure}[H]
\psfig{figure=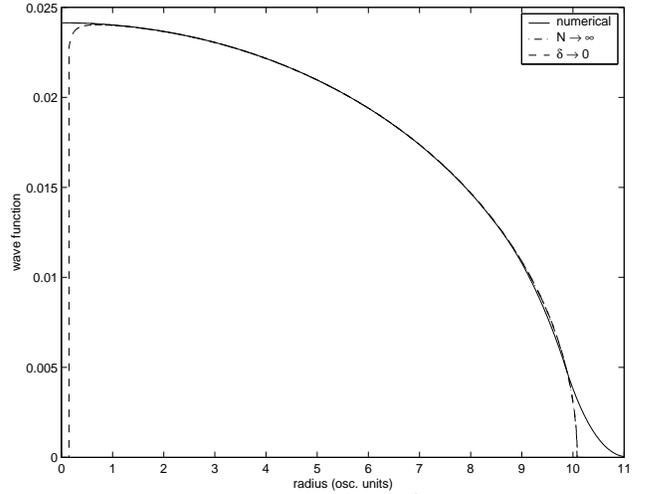,height=6.6cm,width=8.6cm}
\caption[fig:gwf2]{Same as Fig.\ \protect\ref{fig:gwf1} but with
$10^6$ atoms, corresponding to $g_{3} \approx 87,204$.  As the
number of atoms increases, the unphysical core in our zeroth-order
wave function shrinks. Near the origin, the $N \to \infty$ TF and
numerically calculated wave functions overlap, while the $N \to
\infty$ TF and our zeroth-order wave function overlap in the
boundary region. For sufficiently large N, the three wave
functions become indistinguishable.} \label{fig:gwf2}
\end{figure}

We allow the effective dimensionality of the condensate to be a
variable quantity, and we use the parameter $\delta=1/\kappa$ to
scale the GPE in arbitrary dimension, where $\kappa$ depends on
the effective dimensionality of the condensate and on the angular
momentum quantum number.  We have shown that our zeroth-order
($\delta \to 0$) limit of the Gross-Pitaevskii equation provides a
less severe approximation of the kinetic energy than the $N
\rightarrow \infty$ Thomas-Fermi approximation for the ground
state, which neglects the entire kinetic energy. The zeroth-order
($\delta \to 0$) limit is a simple approximation that, in order to
satisfy the minimum uncertainty principle, retains a kinetic
energy contribution, rather than neglecting the entire KE, making
it more accurate and flexible than the ground-state $N \to \infty$
TFA. As shown in Fig.\ \ref{fig:mucompare}, our zeroth-order
approximation is more accurate than the $N \to \infty$ TFA for the
chemical potential. This improvement is caused by the
centrifugal-like term, which brings in the kinetic energy needed
to satisfy the minimum uncertainty principle and which adds a
needed outward push to the wave function. The accuracy of the
zeroth-order approximation is comparable to the lowest order
correction due to the boundary layer at the condensate surface.
Improved accuracy for the ground state is most noticeable for a
moderate number of atoms, the case in which the kinetic energy is
most significant. For a sufficiently large number of atoms with
positive scattering length, the kinetic energy becomes small for
the ground state chemical potential, and the three approximations
converge to the numerical solution of the GPE.

The core near the origin and the presence of the angular momentum
quantum number, $l$, in the scaling parameter, $\delta =1/\kappa$,
make the zeroth-order ($\delta \to 0$) density an especially good
starting point for studying properties of vortices. The
ground-state $N \to \infty$ TFA is unable to accommodate such
states, but it can be extended to include vortices by introducing
the gradient of the phase from the Laplacian
\cite{crescim,fetter,sinha,svid}.  We expect higher order,
finite-$\delta$ corrections to further refine the shape of our
zeroth-order density for the ground state and vortex states.

We have shown that the dimensional scaling formalism is conducive
to analysis of condensates of any dimension.  We outlined how
simple yet accurate approximations can be achieved for any
effective D, and we demonstrated the improved numerical results
for $D=3$. In addition to 3D BEC, the dimensional scaling
formalism provides a useful analytical tool in the study of BEC in
lower effective dimensionality.  In future work, we plan to test
the approximation presented here on other observables and states
of BEC in D-dimensions, and we are extending the methods presented
herein to D-dimensional cylindrical coordinates, where the
anisotropy parameter is included explicitly for treatment of
axially symmetric traps with arbitrary anisotropy.

\vspace{1ex}

\begin{acknowledgments}
This research was supported by the Office of Naval Research Grant
No. N00014-00-1-0576.  We would like to thank Olen Boydstun,
Michael Crescimanno, Martin Dunn, and Brett Esry for helpful
discussions.
\end{acknowledgments}


\begin{thebibliography}{99}
\bibitem{note1} Dimensional-scaling studies have been
applied to a large number of phenomena including the atomic Stark
and diamagnetic Zeeman effects, the hydrogen atom in parallel
electric and magnetic fields, the laser-dressed hydrogen atom;
two-electron and many-electron atoms and ions, molecules, Rydberg
states, theory of chemical bonding, van der Waals coefficients,
atomic clusters, atomic polarizabilities and multipole sum rules,
exotic atoms, electron clusters in traps, quasistationary states,
density functional theory, tunnelling problems, potential
scattering problems, and the computation of the rotational spectra
of molecules. Dimensional-scaling studies have also been applied
to problems in nuclear physics, mesons and baryons, relativistic
quantum systems, quantum field theory, the jellium model, solid
state physics, quantum dots, the calculation of virial
coefficients and statistical physics\cite{high_L1}.
\end{thebibliography}
\end{document}